\documentclass[twocolumn,english,reprint,amsmath,amssymb, onecolumn,nofootinbib,11pt]{revtex4-2}
\usepackage[T1]{fontenc}
\usepackage[latin9]{inputenc}
\makeatletter
\usepackage{float}
\usepackage{physics}
\usepackage{amsfonts}
\usepackage{latexsym}
\usepackage{epsfig}
\usepackage{graphicx}
\usepackage{tikz}
\definecolor{greatblue}{RGB}{40,120,181}
\definecolor{greatred}{RGB}{200,36,35}
\usepackage[colorlinks,linkcolor=greatblue,anchorcolor=blue,citecolor=greatred]{hyperref}

\usepackage{dcolumn}
\usepackage{bm}

\makeatother

\usepackage[english]{babel}

\begin{document}
\preprint{preprintnumbers}{CTP-SCU/2026006}

\title{Exact and Finite de Sitter QFT from CFT}




\author{Haitang Yang}
\email{hyanga@scu.edu.cn}
\affiliation{College of Physics, Sichuan University, Chengdu, 610065, China}


\begin{abstract}

Parallel to the  AdS Scale-Space construction in \cite{Yang:2026AdS},
we construct a $(d+1)$-dimensional  de Sitter QFT directly from $d$-dimensional CFT data. The dS geometry is the moduli space of oriented balls, and the lifted operators are obtained by conformal-family Casimir completion. The Euclidean parent CFT gives a finite wavefunction representation, while the Minkowski parent CFT gives a double-time representation that is unitary in the parent-time polarization. This provides a CFT-based framework for reexamining puzzles of dS and double-time QFT.
\end{abstract}

\maketitle

\tableofcontents

\section{Introduction}
The AdS/CFT correspondence gives a concrete realization of holography in spacetimes with negative cosmological constant \cite{Maldacena:1997re,Gubser:1998bc,Witten:1998qj}. Its standard GKPW form is asymptotic: a bulk field is specified by its boundary behavior, and CFT generating functionals are related to bulk  actions with prescribed boundary data. The dS/CFT correspondence was proposed as a positive cosmological constant analogue of this idea \cite{Strominger:2001pn}. In the original proposal, quantum gravity in de Sitter space is related to a conformal field theory living at spacelike infinity. Closely related discussions emphasized the special difficulties of quantum gravity in de Sitter space, in particular the role of meta-observables, the absence of an ordinary asymptotic S-matrix, and the possibility of a finite-dimensional Hilbert space \cite{Witten:2001kn,Bousso:2000nf}. The geometry and asymptotic conformal structure of de Sitter space were further reviewed in \cite{Spradlin:2001pw,Anninos:2012qw}, and boundary stress tensors for asymptotically de Sitter spaces were developed in \cite{Balasubramanian:2001nb}.

Despite the analogy with AdS/CFT, dS/CFT is not simply AdS/CFT with the sign of the cosmological constant changed. The boundary of global de Sitter is spacelike, not timelike. The putative dual CFT is often Euclidean and may be non-unitary. Massive stable scalar fields may lead to complex conformal weights \cite{Strominger:2001pn}. In the higher-spin realization, Vasiliev theory in de Sitter space is conjecturally dual to a Euclidean $\mathrm{Sp}(N)$ vector model with anticommuting scalars \cite{Anninos:2011ui}. Moreover, the dS/CFT dictionary is naturally formulated in terms of the wavefunction of the universe rather than an ordinary unitary transition amplitude \cite{Maldacena:2002vr,McFadden:2009fg}.

Different from the traditional route, and based on a series of CFT-first constructions 
\cite{Jiang:2025jnk,Jiang:2026juf,Yang:2026mrt}, 
we recently developed a Conformal-Moduli-Space QFT (CMS-QFT) formalism 
\cite{Yang:2026AdS}\footnote{As this preceding work remains on hold by arXiv for unknown reasons, 
we are compelled to include an appendix \ref{appendix} giving a concise overview.}.
In this approach, the CFT is the starting point and no duality is assumed. The bulk-boundary picture is not taken as primitive; it appears only in a singular limit. The relevant input is the triplet consisting of the CFT, the state, and the organization pattern (geometry) of the CFT data.

In \cite{Yang:2026AdS}, an Open-Solid-Torus (OST) geometry assigns an extra degree of freedom to each point of the CFT. In the symmetric configuration this extra degree of freedom behaves as a scale, and the induced Scale-Space turns out to carry an AdS$_{d+1}$ metric. The CFT operators are lifted to this Scale-Space by conformal-family Casimir completion. The connected CFT correlators are then completed, giving a hierarchy of kernels from which one defines the connected generating functional and the 1PI effective action of an exact finite curved QFT. In this construction, no assumptions such as large $N$, strong coupling, or heavy operators are required. The ordinary GKPW relation is recovered only as a singular boundary limit.

For the de Sitter case, however, an important CFT-first precedent is the program developed by de Boer et al.~\cite{deBoer:2015kda,deBoer:2016pqk}. Ref.~\cite{deBoer:2015kda} showed that the moduli space of spherical regions on a fixed Lorentzian time slice carries an auxiliary de Sitter geometry, with the ball radius playing the role of de Sitter time. Ref.~\cite{deBoer:2016pqk} substantially extended this idea by organizing CFT observables on the full moduli space of causal diamonds. The resulting generalized kinematic space is $2d$-dimensional with signature $(d,d)$, and conformal-family observables obey Casimir equations on this space. In particular, these works already established an important reorganization of CFT data on conformal moduli spaces. See also Ref.~\cite{Haehl:2025ehf} for a recent development of the causal-diamond/OPE-block program.

The moduli problem considered here is nevertheless different. Rather than starting from spatial balls on Lorentzian time slices and completing them to causal diamonds, we start directly from full $d$-dimensional balls in a Euclidean CFT. Their centers already carry all $d$ CFT coordinates, while the radius supplies precisely one additional modulus; the resulting complete ball-moduli space is therefore $(d+1)$-dimensional and carries the dS$_{d+1}$ metric derived below. Our subsequent purpose is also different: after constructing this finite moduli space, we transport the complete hierarchy of connected CFT correlators to it and use the resulting kernels as the defining dynamical data of a curved QFT.

The  question addressed in this paper is therefore whether the CMS-QFT logic developed in Ref.~\cite{Yang:2026AdS} can be realized on this de Sitter ball-moduli space. The scalar CFT operators are lifted by conformal-family Casimir completion. Applying the completion to every external leg of every connected CFT correlator defines finite kernels on the ball-moduli space. These kernels define a finite connected generating functional $W_{\mathcal M}[J]$ and a finite 1PI effective action $\Gamma_{\mathcal M}[\Phi]$. Thus every CFT defines a finite Conformal-Moduli-Space QFT with de Sitter geometry, whose dynamics are inherited from the parent CFT correlators.

The construction can start from two parents: a Euclidean CFT or a Minkowski CFT. 
If the parent is a Euclidean CFT, one obtains a single-time de Sitter ball-moduli space. This gives a finite version of the usual dS/CFT wavefunction representation \cite{Strominger:2001pn,Witten:2001kn,Maldacena:2002vr,McFadden:2009fg}. Since the parent theory is Euclidean, this representation does not by itself provide a Lorentzian Hilbert space or a preferred Hamiltonian time. The familiar features of dS/CFT, such as state selection ambiguity, contour dependence, and the wavefunction character of the dictionary, should therefore be understood as features of this single-time representation rather than as defects of the finite Conformal-Moduli-Space construction.

If the parent is instead a Minkowski CFT, the CFT time $t$ remains present after the lift. The resulting geometry has two timelike directions. This appears unusual from the viewpoint of ordinary double-time field theory. However, the present construction is not obtained by canonically quantizing a local field theory with two independent Hamiltonian times. It is obtained by constructing $T$-dependent completed operators on the original Minkowski CFT Hilbert space. The physical Hamiltonian is the parent Hamiltonian $H_t$, and all completed correlators are evaluated in the $H_t$-vacuum. In this precise sense, the double-time QFT is parent-time unitary.

Now that a Minkowski CFT can induce both an AdS QFT via OST and a double-time dS QFT via ball-moduli,
it stands to reason that additional Conformal-Moduli-Spaces exist 
which yield their respective curved QFT counterparts. 
These CMS-QFTs  form   a ``duality'' web through the underlying CFT.

The paper is organized as follows. Section II derives the ball-moduli geometry from the ambient conformal embedding. Section III constructs the scalar Casimir completion. Section IV develops the Euclidean CFT realization and its relation to the traditional dS wavefunction dictionary. Section V develops the Minkowski CFT realization and explains in what sense the double-time theory is parent-time unitary. Section VI compares the two representations and discusses de Sitter non-unitarity, vacuum ambiguity, observer reductions, infrared issues, and the relation to AdS. Section VII discusses 
the multiplicity of CMS presentations, an interpretation of dualities, and selection of physical time. 
Section VIII summarizes the construction and provides further discussion. 

\section{Geometry of the Ball-Moduli Space}

We will first work on Euclidean CFT and turn to Minkowski in Section \ref{sec:Min}.
The de Sitter geometry associated with moduli of spherical regions has important precedents in Refs.~\cite{deBoer:2015kda,deBoer:2016pqk}. Here the finite object is a full $d$-dimensional Euclidean ball, and we use the standard ambient embedding of conformal geometry to derive its $(d+1)$-dimensional moduli metric in a form suited to the subsequent QFT construction.

Let the original Euclidean CFT live on $\mathbb R^d$. For each center $X^\mu$ and radius $T>0$, define the open ball
\begin{equation}
B(X,T)=\{x:|x-X|^2<T^2\}.
\label{eq:balls}
\end{equation}
Its boundary sphere is
\begin{equation}
S(X,T)=\{x:|x-X|^2=T^2\}.
\end{equation}
The orientation of the sphere is part of the data. In this paper, unless stated otherwise, 
the orientation is chosen so that the inside of the open ball is positive.

The distinction between the open ball and the sphere is important. 
The open ball is the  CFT region. The sphere is the boundary  of that region. 
Nevertheless,
the $(d+1)$-dimensional moduli space $\mathcal M$, with coordinates $Y=(X^\mu,T)$, 
may be described either as the space of inside-oriented balls or as the space of their oriented boundary spheres.

Introduce the ambient space $\mathbb R^{d+1,1}$ with inner product
\begin{equation}
U\cdot V
=
-U^{-1}V^{-1}+U^\mu V^\mu+U^{d+1}V^{d+1}.
\label{eq:ambient-inner}
\end{equation}
A CFT point $x^\mu\in\mathbb R^d$ is represented by the null vector
\begin{equation}
P^A(x)
=
\left(
\frac{1+x^2}{2},
x^\mu,
\frac{1-x^2}{2}
\right), \qquad P(x)^2=0,
\label{eq:ambient-map}
\end{equation}
where the Poincar\'e gauge is chosen
\begin{equation}
P^{-1} +P^{d+1} =1.
\label{eq:ambient-Poincare}
\end{equation}
Thus the original CFT space is represented as the projective null cone, and the Euclidean conformal group acts linearly on the ambient coordinates.

Let $V^A$ be a non-null ambient vector. 
Since 
\begin{equation}
V\cdot P(x)
=
-\frac{V^{-1}+V^{d+1}}{2}x^2
+
V_\mu x^\mu
+
\frac{-V^{-1}+V^{d+1}}{2} \equiv a x^2 + b_\mu x^\mu + c,
\end{equation}
the incidence relation
\begin{equation}
V\cdot P(x)=0,
\end{equation}
defines a sphere  or a plane. The sign of $V\cdot P(x)$ selects one side of the sphere. 
In order to match  the ball $B(X,T)$ in Eq. (\ref{eq:balls}), 
we choose the normalization and orientation by imposing
\begin{equation}
V(X,T)\cdot P(x)
=
\frac{T^2-|x-X|^2}{2T}.
\end{equation}
Then
\begin{equation}
V(X,T)\cdot P(x)=0
\end{equation}
is precisely the boundary sphere, and
\begin{equation}
V(X,T)\cdot P(x)>0
\end{equation}
is precisely the inside of the ball.

Solving for $V(X,T)$, one obtains the normalized vector
\begin{equation}
V= \left(\frac{1+X^2-T^2}{2T}, \frac{X^\mu}{T}, \frac{1-X^2+T^2}{2T} \right), \qquad V^2=1.
\label{eq:ambient-norm}
\end{equation}
The metric on the ball-moduli space  $\mathcal M$ is induced from the ambient metric:
\begin{equation}
\boxed{ds_{\mathcal M}^2
= dV\cdot dV
=\frac{-dT^2+dX^\mu dX^\mu}{T^2}.}
\label{eq:dS-metric}
\end{equation}
This is the Poincare patch metric of $\mathrm{dS}_{d+1}$. However, in order to avoid confusion with the usual bulk-boundary interpretation, we will mostly refer to it as the ball-moduli space $\mathcal M$.

The invariant of two oriented balls is the ambient inner product
\begin{equation}
\mathcal Z_{12}
=
V(X_1,T_1)\cdot V(X_2,T_2).
\end{equation}
Using the explicit expression for $V(X,T)$, one obtains
\begin{equation}
\mathcal Z_{12}
=
\frac{
T_1^2+T_2^2-|X_1-X_2|^2
}
{2T_1T_2}.
\end{equation}
This is the signed inversive invariant of two oriented balls.
The sign and range of $\mathcal Z_{12}$ have geometric meaning, 
\begin{eqnarray}
\mathcal Z_{12}>1 &\qquad& \hbox{Nested balls}, \nonumber\\
\mathcal Z_{12}=1 &\qquad& \hbox{Internal tangency}, \nonumber\\
1> \mathcal Z_{12}>-1 &\qquad& \hbox{Intersecting boundary spheres},\\
\mathcal Z_{12}=-1 &\qquad& \hbox{external tangency}, \nonumber\\
\mathcal Z_{12}<-1 &\qquad& \hbox{externally separated balls}\nonumber.
\end{eqnarray}

\section{Casimir Completion of Scalar Conformal Families}

In this section, following the same pattern as in the AdS Scale-Space construction \cite{Yang:2026AdS}, we lift scalar operators of the $d$-dimensional CFT$_d$ to the induced $(d+1)$-dimensional ball-moduli space $\mathcal M$. The construction of conformal-family observables obeying Casimir equations on conformal moduli spaces is already central to Ref.~\cite{deBoer:2016pqk}. Related representation-theoretic constructions include OPE blocks and bulk-local-state/descendant completions \cite{Nakayama:2015mva,Wang:2015qfa,Czech:2016xec,Doi:2024nty}. These works provide important precedents for the single-family kinematic step. The present lift is a differential Casimir completion on the full-ball de Sitter moduli space; dynamics enters later through the complete hierarchy of connected CFT correlators.

Let $O_\Delta(X)$ be a scalar primary of dimension $\Delta$. The corresponding primary state satisfies
\begin{equation}
K_\mu|\Delta\rangle=0, \quad
D|\Delta\rangle=\Delta|\Delta\rangle,\quad
M_{\mu\nu}|\Delta\rangle=0.
\end{equation}
The quadratic conformal Casimir has eigenvalue
\begin{equation}
C_2(\Delta)=\Delta(\Delta-d).
\end{equation}

At a canonical point of the ball-moduli space, the lifted scalar must be invariant under the rotational stabilizer. Therefore the scalar descendants are generated by powers of $P^2$. A general rotationally invariant descendant state has the form
\begin{equation}
|O_\Delta\rangle_{\rm c}
=
\sum_{n=0}^{\infty}
c_n (P^2)^n|\Delta\rangle.
\label{eq:canonical}
\end{equation}
General scale $T$ and center $X$ are achieved by acting $e^{X\cdot P}e^{(\log T)D}$ on the canonical point, which yields
\begin{equation}
O_\Delta^{\mathcal M}(X,T)
=
T^\Delta
\sum_{n=0}^{\infty}
c_n T^{2n}\Box_X^n O_\Delta(X).
\label{eq:ansatz}
\end{equation}
The scalar Laplacian   associated with the metric (\ref{eq:dS-metric}) is
\begin{equation}
\nabla_{\mathcal M}^2
=
T^2(\Box_X-\partial_T^2)
+
(d-1)T\partial_T.
\end{equation}
The Casimir equation for the completed scalar is
\begin{equation}
\left(
\nabla_{\mathcal M}^2 - \Delta(d-\Delta) \right)
O_\Delta^{\mathcal M}(X,T)=0.
\end{equation}
This is not assumed as a bulk wave equation. 
It is the coordinate realization of the conformal Casimir on the scalar family.
If the same equation is interpreted as a conventional de Sitter bulk wave equation, it corresponds to
\begin{equation}
m^2 = \Delta(d-\Delta).
\end{equation}
In that conventional interpretation, principal-series fields with sufficiently large mass are described by complex weights. In the present construction, however, $\Delta$ is the scaling dimension of the parent CFT operator and the equation is a Casimir constraint, not an assumed bulk field equation.

Substituting Eq. (\ref{eq:ansatz}) into the Casimir equation gives the recursion
\begin{equation}
c_{n-1}
-
4n\left(\Delta-\frac d2+n\right)c_n
=
0,
\qquad
n\geq1.
\end{equation}
With normalization $c_0=1$, this gives
\begin{equation}
c_n
=
\frac{1}
{
4^n n!
\left(\Delta-\frac d2+1\right)_n
},
\end{equation}
where $(a)_n\equiv a(a+1)\cdots (a+n-1)$ and $(a)_0 =1$.
Therefore, we get the lifted CFT operators in the ball-moduli space,
\begin{equation}
\boxed{
O_\Delta^{\mathcal M}(X,T)
=
T^\Delta
{}_0F_1
\left(
\Delta-\frac d2+1;
\frac{T^2}{4}\Box_X
\right)
O_\Delta(X).}
\label{eq:Operator-Completion}
\end{equation}

In the Open Solid Torus (OST) Scale-Space construction given in \cite{Yang:2026AdS}, 
the completed scalar family has the form
\begin{equation}
O_\Delta^{\rm SS}(X,Z)
=
Z^\Delta
{}_0F_1
\left(
\Delta-\frac d2+1;
-\frac{Z^2}{4}\Box_X
\right)
O_\Delta(X).
\end{equation}
The sign difference is fixed by the signature of the scale direction. 
In OST Scale-Space the scale coordinate is spacelike, while in the ball-moduli space the radius coordinate is timelike.

\section{Euclidean Parent: Single-Time de Sitter and Wavefunction Representation}

Let the Euclidean CFT have connected scalar correlators
\begin{equation}
G^{E}_{a_1\cdots a_n,c}(X_1,\ldots,X_n)
=
\left\langle
O_{a_1}(X_1)\cdots O_{a_n}(X_n)
\right\rangle_{E,c}.
\end{equation}
For each operator $O_a$ of dimension $\Delta_a$, define the completion operator
\begin{equation}
\boxed{
R_{\Delta_a}^{E}(X,T)
=
T^{\Delta_a}
{}_0F_1
\left(
\Delta_a-\frac d2+1;
\frac{T^2}{4}\Box_X
\right).}
\end{equation}
Using the operator completion (\ref{eq:Operator-Completion}),  
the completed connected kernels on ball-moduli space  are
\begin{equation}
\boxed{
G^{E-\mathcal{M}}_{a_1\cdots a_n,c}(Y_1,\ldots,Y_n)
=
\prod_{i=1}^{n}
R_{\Delta_{a_i}}^{E}(X_i,T_i)
\,
G^{E}_{a_1\cdots a_n,c}(X_1,\ldots,X_n),}
\end{equation}
where
\begin{equation}
Y_i=(X_i,T_i).
\end{equation}

This definition lifts the full CFT data. OPE coefficients, conformal blocks, and crossing constraints enter through the original CFT correlators. The moduli-space kernels are finite objects derived from CFT.

The invariant measure on the ball-moduli space is
\begin{equation}
d\mu(Y)
=
\frac{d^dX\,dT}{T^{d+1}}.
\end{equation}
Introduce sources $J^a(Y)$. The finite connected generating functional is
\begin{equation}
W_{E-\mathcal M}[J]
=
\sum_{n=1}^{\infty}
\frac{1}{n!}
\int
\prod_{i=1}^{n}
d\mu(Y_i)
J^{a_i}(Y_i)
G^{E-\mathcal M}_{a_1\cdots a_n,c}(Y_1,\ldots,Y_n).
\end{equation}
The response field is
\begin{equation}
\Phi_a(Y)
=
\frac{\delta W_{E-\mathcal M}[J]}{\delta J^a(Y)}.
\end{equation}
The finite 1PI effective action is
\begin{equation}
\Gamma_{E-\mathcal M}[\Phi]
=
\int d\mu(Y)\,J^a(Y)\Phi_a(Y)
-
W_{E-\mathcal M}[J].
\end{equation}
The construction is therefore
\begin{equation}
G^E_{n,c}
\longrightarrow
G^{E-\mathcal M}_{n,c}
\longrightarrow
W_{E-\mathcal M}[J]
\longrightarrow
\Gamma_{E-\mathcal M}[\Phi].
\end{equation}
No bulk path integral or bare moduli space action is required to define these objects. A traditional local action on the ball-moduli space, if it exists, is an additional representation of the same completed kernel hierarchy.

Near $T=0$, the completed operator (\ref{eq:Operator-Completion}) behaves as
\begin{equation}
O_\Delta^{E-\mathcal M}(X,T)
=
T^{\Delta}O(X)+O(T^{\Delta+2}).
\end{equation}
Choose a source localized at $T=\epsilon$,
\begin{equation}
J_\epsilon^a(X,T)
=
T^{d+1-\Delta_a}j^a(X)\delta(T-\epsilon).
\end{equation}
Then
\begin{equation}
\int d\mu(Y)
J_\epsilon^a(Y)O_a^{E-\mathcal M}(Y)
=
\int d^dX\,j^a(X)O_a(X)+O(\epsilon^2).
\end{equation}
Therefore
\begin{equation}
\lim_{\epsilon\to0}
W_{E-\mathcal M}[J_\epsilon[j]]
=
W_E[j].
\end{equation}

This is the finite-parent origin of the usual dS/CFT boundary relation. In a semiclassical gravitational phase, one may write schematically
\begin{equation}
Z_E[j]\sim \Psi_{dS}[j].
\end{equation}
In the present formulation, this is not the definition of the theory. It is a singular boundary representation of the finite moduli-space functional.

The Euclidean parent CFT naturally gives a wavefunction representation. This is compatible with the traditional dS/CFT picture, where a Euclidean CFT is associated with the spacelike boundary of de Sitter space \cite{Strominger:2001pn,Witten:2001kn}. In this representation, the timelike direction of the de Sitter moduli space is generated by the radius. Since the parent theory is Euclidean, this representation does not by itself provide a Lorentzian Hilbert space or a preferred Hamiltonian time. The familiar features of dS/CFT, including state-selection ambiguity, contour dependence, and the wavefunction character of the dictionary, should therefore be understood as features of this single-time representation rather than as defects of the  moduli-space construction. This is compatible with the fact that many concrete dS/CFT examples involve non-unitary or ghost-like Euclidean CFTs, such as the $\mathrm{Sp}(N)$ vector model with anticommuting scalars \cite{Anninos:2011ui,Ng:2012xp}.

\section{Lorentzian Parent: Double-Time de Sitter and Parent-Time Unitarity \label{sec:Min}}

\subsection{Well-Defined Double-Time de Sitter QFT}
To obtain a representation with a manifest Hamiltonian time, we now consider a Lorentzian parent CFT, with coordinates and metric
\begin{equation}
x^\alpha=(t,\vec X),\qquad
dx^\alpha dx_\alpha=-dt^2+d\vec X^2.
\end{equation}
The corresponding d'Alembertian is
\begin{equation}
\Box_M=-\partial_t^2+\nabla_{\vec X}^2.
\end{equation}

The key difference from the Euclidean parent is that the Minkowski CFT already has a physical time $t$, 
a Hilbert space $\mathcal H_M$, and a Hamiltonian $H_t$. For a unitary Minkowski CFT,
\begin{equation}
U(t)=e^{-iH_t t}
\end{equation}
is unitary on $\mathcal H_M$. A vacuum state is defined by
\begin{equation}
H_t|0_M\rangle=0.
\end{equation}

The Lorentzian representation of the ball-moduli construction is obtained by replacing the Euclidean quadratic form by the Minkowski one. Equivalently, one may impose the Lorentzian incidence relation
\begin{equation}
V(x,T)\cdot P(y)=\frac{T^2-(y-x)^2_M}{2T},
\end{equation}
where $(y-x)^2_M=-(y^0-t)^2+(\vec y-\vec X)^2$. The induced metric is then
\begin{equation}
\boxed{
ds_{\rm dt}^2
=
\frac{-dT^2+dx^\alpha dx_\alpha}{T^2}
=
\frac{-dT^2-dt^2+d\vec X^2}{T^2}.}
\end{equation}
Thus the induced geometry has two timelike directions. However, this does not mean that we quantize a local field theory with two independent Hamiltonian times. The Hilbert space is entirely inherited from the parent Minkowski CFT. The physical Hamiltonian is $H_t$. The $T$-coordinate labels the conformal family completion.

For a scalar primary $O_\Delta(x)$ in the Minkowski CFT,
the Lorentzian parent analogue of the Euclidean ball-moduli completion (\ref{eq:Operator-Completion}) is
\begin{equation}
\boxed{
O_\Delta^{\rm dt}(x,T)
= R^{\rm dt}_{\Delta}(x,T) \,O_\Delta(x)
\equiv 
T^\Delta
{}_0F_1
\left(
\Delta-\frac d2+1;
\frac{T^2}{4}\Box_M
\right)
O_\Delta(x),
}
\end{equation}
where we defined
\begin{equation}
\boxed{
R^{\rm dt}_{\Delta}(x,T)
=
T^{\Delta}
{}_0F_1
\left(
\Delta-\frac d2+1;
\frac{T^2}{4}\Box_M
\right).}
\end{equation}
The operator $O_\Delta^{\rm dt}(x,T)$ acts on the same Hilbert space as $O_\Delta(x)$:
\begin{equation}
O_\Delta^{\rm dt}(x,T):\mathcal H_M\rightarrow\mathcal H_M.
\end{equation}
The $T$-dependence is generated by descendant completion, not by an independent canonical time evolution.

Let
\begin{equation}
G^M_{a_1\cdots a_N,c}(x_1,\ldots,x_N)
=
\langle 0_M|
O_{a_1}(x_1)\cdots O_{a_N}(x_N)
|0_M\rangle_c
\end{equation}
be connected Minkowski CFT correlators in the $H_t$-vacuum. 
The completed double-time kernels are then
\begin{equation}
\boxed{
G^{\rm dt}_{a_1\cdots a_N,c}(Y_1,\ldots,Y_N)
=
\prod_{i=1}^{N}
R^{\rm dt}_{\Delta_{a_i}}(x_i,T_i)
\,
G^M_{a_1\cdots a_N,c}(x_1,\ldots,x_N),}
\end{equation}
where
\begin{equation}
Y_i=(x_i,T_i) = (t_i,\vec X_i, T_i).
\end{equation}
This equation is the central point of the Lorentzian realization. All completed correlators are defined in a well-defined parent Hilbert space and in a well-defined parent vacuum. The apparent double-time geometry is a geometry of completed correlators. It is not a conventional two-time canonical quantization.

With the double-time invariant measure
\begin{equation}
d\mu_{\rm dt}(Y)
=
\frac{d^dx\,dT}{T^{d+1}}
= \frac{dt\,d^{d-1}\vec X\,dT}{T^{d+1}},
\end{equation}
define the generating functional of connected correlators in the moduli space
\begin{equation}
W_{\rm dt}[J]
=
\sum_{N=1}^{\infty}
\frac{1}{N!}
\int
\prod_{i=1}^{N}
d\mu_{\rm dt}(Y_i)
J^{a_i}(Y_i)
G^{\rm dt}_{a_1\cdots a_N,c}(Y_1,\ldots,Y_N).
\end{equation}
The double-time response field is
\begin{equation}
\Phi_a(Y)
=
\frac{\delta W_{\rm dt}[J]}{\delta J^a(Y)}.
\end{equation}
The 1PI generating functional is
\begin{equation}
\Gamma_{\rm dt}[\Phi]
=
\int d\mu_{\rm dt}(Y)\,J^a(Y)\Phi_a(Y)
-
W_{\rm dt}[J].
\end{equation}

The theory is well-defined in the following sense. If the parent Minkowski CFT is   unitary, 
then $\mathcal H_M$, $H_t$, and $|0_M\rangle$ are well-defined. 
The completed operators are constructed algebraically from conformal 
descendants of operators acting on $\mathcal H_M$. 
Therefore the completed kernels and the functionals $W_{\rm dt}$ and $\Gamma_{\rm dt}$ 
are finite parent-time objects.

\subsection{Distinction from ordinary double-time QFT}

The metric
\begin{equation}
ds_{\rm dt}^2
=
\frac{-dT^2-dt^2+d\vec X^2}{T^2}
\end{equation}
has two negative directions. In ordinary double-time field theory, 
this would raise immediate questions about ghosts, causality, and unitarity. 
In known approaches to double-time physics, 
additional gauge symmetries or constraints are often required to remove unphysical degrees of freedom \cite{Bars:1998cs,Bars:2000qm,Bars:2010xi}.

The present construction has a different status. 
It does not begin with an autonomous local field theory on a double-time manifold. 
It begins with a unitary Minkowski CFT. 
The double-time coordinate $T$ is introduced through conformal family completion. 
The Hilbert space is not enlarged by a second canonical time evolution. Thus
\begin{equation}
\mathcal H_{\rm dt}=\mathcal H_M,
\end{equation}
and
\begin{equation}
H_{\rm phys}=H_t.
\end{equation}
The $T$-direction is timelike in the moduli-space metric, 
but it is not the Hamiltonian time used to define the Hilbert-space inner product. 
It is a moduli-time direction appearing in the Casimir equation.

In this precise sense, the double-time construction is parent-time unitary. 
Its unitarity is inherited from the Minkowski CFT, 
not established by independently quantizing a double-time local action.

\subsection{Relation to AdS Scale-Space}

The double-time construction resembles analytic continuation in some formulas, but it should not be identified with an ordinary Wick rotation of AdS. In AdS Scale-Space with a Lorentzian parent CFT, the scalar completion has the schematic form
\begin{equation}
O_\Delta^{\rm SS}(x,Z)
=
Z^\Delta
{}_0F_1
\left(
\Delta-\frac d2+1;
-\frac{Z^2}{4}\Box_M
\right)
O_\Delta(x).
\label{eq:AdS O}
\end{equation}
The double-time completion is instead
\begin{equation}
O_\Delta^{\rm dt}(x,T)
=
T^\Delta
{}_0F_1
\left(
\Delta-\frac d2+1;
\frac{T^2}{4}\Box_M
\right)
O_\Delta(x).
\end{equation}
At the level of the scalar descendant kernel, the two completions are formally related by the analytic replacement of the scale-squared variable. This relation should not be interpreted as an identification of the AdS scale coordinate $Z$ with the ball radius $T$. The two variables arise from different   CFT moduli problems: $Z$ is the scale coordinate of the point-pair construction, while $T$ is the radius coordinate of an oriented sphere or ball.

The present route is therefore
\begin{equation}
\text{unitary Minkowski CFT}
\longrightarrow
\text{descendant completion}
\longrightarrow
\text{double-time moduli kernels}.
\end{equation}
The parent time $t$ is not Wick-rotated. The parent Hilbert space is not analytically continued. The inner product remains the Minkowski CFT inner product.

\section{Two Representations and de Sitter Puzzles}

The construction gives two related but distinct representations:
\begin{equation}
\text{Euclidean CFT}
\longrightarrow
\text{single-time de Sitter moduli space}
\longrightarrow
\text{wavefunction rep.},
\end{equation}
and
\begin{equation}
\text{Minkowski CFT}
\longrightarrow
\text{double-time de Sitter moduli space}
\longrightarrow
\text{parent-time unitary rep.}.
\end{equation}

The first representation is close to traditional dS/CFT. It is naturally Euclidean and wavefunction-like. 
The second representation keeps a Lorentzian parent Hilbert space and a physical Hamiltonian $H_t$. 
Therefore it gives a different interpretation of the same finite moduli-space idea.

This distinction suggests that some familiar dS difficulties  may be features of the Euclidean or 
single-time representation.

\subsection{Non-unitarity}

Traditional dS/CFT often involves a Euclidean boundary theory, and concrete higher 
spin examples involve non-unitary or ghost-like CFTs \cite{Anninos:2011ui}. 
From the present viewpoint, this non-unitarity is a feature of the Euclidean wavefunction representation. 
The Euclidean construction has no parent Hamiltonian time. It computes a wavefunction-type object.

By contrast, the Minkowski CFT construction begins with a unitary Lorentzian theory. 
If the parent CFT is unitary, then all completed correlators are evaluated in a unitary Hilbert space. 
The non-unitarity of the Euclidean representation is therefore not inherited 
by the Lorentzian parent representation.

This does not mean that the double-time geometry is an ordinary double-time 
unitary QFT in the canonical sense. It indicates that the moduli-space
theory is unitary in the parent-time polarization.

\subsection{Vacuum ambiguity}

In ordinary de Sitter physics, a preferred global Hamiltonian vacuum is absent. This is tied to the absence of a globally preferred timelike Killing vector and to the fact that de Sitter time evolution is more naturally a cosmological or wavefunction evolution than a standard Hamiltonian evolution. Useful state prescriptions exist, most notably the Bunch-Davies or Euclidean vacuum \cite{Chernikov:1968zm,Bunch:1978yq}, but symmetry alone does not define a unique Hamiltonian ground state; the family of de Sitter invariant $\alpha$-vacua and their interacting pathologies illustrate the subtlety \cite{Mottola:1984ar,Allen:1985ux,Collins:2003zv}.

In the Conformal-Moduli-Space construction, the Euclidean representation inherits this issue. 
The $T$-direction is the only time-like direction of the moduli space, 
and the corresponding boundary object is naturally a wavefunction. 
State selection appears as a choice of contour, real section, or $i0$ prescription.

In the Minkowski CFT representation, the situation is different. 
The parent time $t$ remains a Killing direction of the double-time metric, 
and the parent Hamiltonian $H_t$ defines a vacuum:
\begin{equation}
H_t|0_M\rangle=0.
\end{equation}
The completed kernels are then
\begin{equation}
G_n^{\rm dt}
=
\langle 0_M|
O_1^{\rm dt}\cdots O_n^{\rm dt}
|0_M\rangle_c.
\end{equation}
Thus the question of a de Sitter $T$-vacuum is replaced by the question of 
how an $H_t$-vacuum correlator theory is represented in the $T$-moduli direction.

This reframes the vacuum problem. It does not prove that every single-time 
dS vacuum ambiguity disappears. It shows that a well-defined parent vacuum can 
exist before one reduces to a single-time dS representation.

%
%
%
%
%
%
%

\section{Multiplicity of CMS Presentations, Dualities and Physical Time Polarizations}

In \cite{Yang:2026AdS}, 
we showed how to use Open-Solid-Torus geometry to generate an AdS QFT from a CFT.
Together with the (double-time) dS construction above, the CMS framework suggests 
an instructive separation between CFT data, its finite presentations, and their possible spacetime interpretations. Let 
\begin{equation}
\mathcal{D}_{\mathcal{C}}
=
\left\{
\mathcal{A}_{\mathcal{C}},\;
\langle O_{a_1}\cdots O_{a_n}\rangle,\;
C_{abc},\;
\Delta_a,\;
\ell_a,\;
\text{Ward identities},\ldots
\right\}
\label{eq:full-conformal-data}
\end{equation}
denote the full data of the underlying CFT  $\mathcal{C}$.
Schematically, for a CMS $\mathcal{M}$, we write
\begin{equation}
\mathcal{U}_{\mathcal{M}}:\mathcal{D}_{\mathcal{C}}\longrightarrow \mathcal{Q}[\mathcal{M}],
\label{eq:cms-lift-map-U}
\end{equation}
where $\mathcal{Q}[\mathcal{M}]$ is not an independent theory, 
but a finite presentation of the same $\mathrm{CFT}$ data. 

The essential point is that the algebraic structure $\mathcal{A}_{\mathcal{M}}$ 
on $\mathcal{Q}[\mathcal{M}]$ is inherited from the $\mathrm{CFT}$ algebra, 
\begin{equation}
\mathcal{A}_{\mathcal{M}}
=
\mathcal{U}_{\mathcal{M}}\bigl(\mathcal{A}_{\mathcal{C}}\bigr)
\label{eq:cms-induced-algebra}
\end{equation}
with all products, commutators, OPE data, and Ward identities transported through the lift. 
Thus the CMS presentation does not introduce a new microscopic algebra; 
it {\it reorganizes} the original one in a finite moduli space basis.
The same statement holds for correlation functions, 
the source functional and the corresponding 1PI functional.

Usually a CMS admits an inverse  map
\begin{equation}
\mathcal{U}^{-1}_{\mathcal{M}}:\mathcal{Q}[\mathcal{M}]\longrightarrow \mathcal{D}_{\mathcal{C}},
\qquad
\mathcal{U}^{-1}_{\mathcal{M}}\,\mathcal{U}_{\mathcal{M}}=\mathrm{Id}.
\label{eq:cms-inverse-reconstruction-V}
\end{equation}
Thus two admissible CMS presentations $\mathcal{M}$ and $\mathcal{N}$ are related by the finite presentation map
\begin{equation}
\mathcal{T}_{\mathcal{M}\to\mathcal{N}}
=
\mathcal{U}_{\mathcal{N}}\mathcal{U}^{-1}_{\mathcal{M}}.
\label{eq:cms-presentation-transition-map}
\end{equation}
These maps obey the composition law
\begin{equation}
\mathcal{T}_{\mathcal{N}\to\mathcal{P}}
\mathcal{T}_{\mathcal{M}\to\mathcal{N}}
=
\mathcal{T}_{\mathcal{M}\to\mathcal{P}},
\qquad
\mathcal{T}_{\mathcal{M}\to\mathcal{M}}
=
\mathrm{Id}.
\label{eq:transition-map-composition-identity}
\end{equation}

Therefore, admissible CMS presentations form a  group.
The group elements form a ``duality'' web.
The  ``dualities'' are not independent dynamical postulates; 
they are induced by changing the finite presentation of $\mathcal{D}_{\mathcal{C}}$.


A CMS may carry a natural metric.
One lesson from the double-time dS construction above is that
a physical spacetime interpretation requires an additional polarization. In particular, the presence of a Lorentzian signature in a CMS metric does not automatically identify the corresponding CMS coordinate as physical time. A physical time interpretation requires data of the form
\begin{equation}
\left(
\mathcal{H},\;
H_{\rm phys},\;
\tau,\;
\text{positivity},\;
\text{causality/locality conditions}
\right),
\label{eq:physical-polarization-data-set}
\end{equation}
where $\tau$ is the chosen physical time parameter and
\begin{equation}
\frac{\mathrm{d}}{\mathrm{d}\tau}O(\tau)
=
i\bigl[H_{\rm phys},O(\tau)\bigr]
\label{eq:physical-time-evolution-commutator}
\end{equation}
defines the physical evolution. This time polarization is not determined by the CMS metric alone. It is an additional interpretive structure placed on top of the finite CMS presentation.

The resulting hierarchy is therefore as follows. 
First, there is an invariant layer, namely the CFT data $\mathcal{D}_{\mathcal{C}}$. 
Second, there is a presentation layer, consisting of admissible CMS representations 
$\mathcal{Q}[\mathcal{M}]$, with their induced algebra, source functional, effective action, and kinematic geometries. Third, there is a physical-polarization layer, where one selects a Hilbert-space interpretation, a physical time, and possibly an effective spacetime geometry for a particular state or sector.

Thus the multiplicity of admissible CMS choices should not be read as multiple independent physical theories. Rather, it expresses the fact that the same CFT data can admit many finite geometric organizations. Physical spacetime, when present, is a further polarization or effective interpretation of such organized data.

\section{Summary and Discussion}

We have constructed exact and finite de Sitter  QFT directly from CFT data. 
The geometric starting point is the moduli space of oriented balls. 
By representing boundary spheres as ambient hyperplane sections of the projective null cone, 
we showed that the  moduli space of balls carries de Sitter geometry.
With conformal family completion, the operators and subsequently the connected correlators of CFT
are lifted   to the ball-moduli space. The generating functionals of connected correlators and 1PI
are thus defined in the ball-moduli space.
The usual dS wavefunction dictionary is recovered only as a singular boundary representation.

There are two realizations, distinguished by the starting parent CFT:  Euclidean CFT or  Minkowski CFT.
The Euclidean parent CFT gives the traditional single-time de Sitter wavefunction picture. 
The Lorentzian parent CFT gives a double-time moduli space, 
but it also keeps the original parent Hilbert space and the parent Hamiltonian $H_t$. 
Therefore its completed correlators are parent-time unitary. 
The apparent double-time geometry is not an autonomous double-time canonical quantization; 
it is a moduli space geometry of completed correlators inside a unitary Minkowski CFT.

This distinction suggests a new organization of de Sitter holography. The Euclidean representation explains why traditional dS/CFT naturally looks wavefunction-like and may be non-unitary in concrete examples. The Lorentzian representation suggests that a unitary parent theory may underlie the same finite moduli space structure. Understanding the precise reduction from the parent-time unitary representation to single-time de Sitter observables is an important open problem.

The double-time realization also suggests a possible reinterpretation of cosmological de Sitter physics. The time used in de Sitter cosmology need not be the microscopic Hamiltonian time. The parent Minkowski CFT supplies a unitary time $t$, while the de Sitter variable $T$ is a moduli or scale-time. Many familiar de Sitter vacuum ambiguities may therefore reflect an attempt to quantize with respect to an effective cosmological time rather than the parent Hamiltonian time. After a reduction $T=T(t)$, however, the induced single-time cosmology generally need not preserve the parent $t$-translation symmetry. A real cosmological prediction therefore requires an additional principle selecting the reduction and specifying how matter clocks couple to the induced geometry.

Finally, since both AdS and double-time dS are different organizations of the same CFT data, 
they must be dual in the traditional sense. It would be of importance to build the specific
dictionary between these two theories.

\vspace*{3.0ex}
\begin{acknowledgments}
\paragraph*{Acknowledgments.} 
This work is supported by NSFC (Grant No. 12275184).
\end{acknowledgments}

\bibliographystyle{unsrturl}
\bibliography{ref202606,dS-QFT-V3-added-refs}

\begin{thebibliography}{10}

\bibitem{Yang:2026AdS}
Haitang Yang.
\newblock Exact and finite ads qft from cft.
\newblock 2026.
\newblock Preprint; precursor of the present work, under arXiv moderation at
  the time of submission. The manuscript is available at
  https://scipost.org/submissions/scipost\_202607\_00082v1.

\bibitem{Maldacena:1997re}
Juan~Martin Maldacena.
\newblock {The Large N limit of superconformal field theories and
  supergravity}.
\newblock {\em Adv. Theor. Math. Phys.}, 2:231--252, 1998.
\newblock \href {http://arxiv.org/abs/hep-th/9711200}
  {\path{arXiv:hep-th/9711200}}, \href
  {http://dx.doi.org/10.1023/A:1026654312961}
  {\path{doi:10.1023/A:1026654312961}}.

\bibitem{Gubser:1998bc}
S.~S. Gubser, Igor~R. Klebanov, and Alexander~M. Polyakov.
\newblock {Gauge theory correlators from noncritical string theory}.
\newblock {\em Phys. Lett. B}, 428:105--114, 1998.
\newblock \href {http://arxiv.org/abs/hep-th/9802109}
  {\path{arXiv:hep-th/9802109}}, \href
  {http://dx.doi.org/10.1016/S0370-2693(98)00377-3}
  {\path{doi:10.1016/S0370-2693(98)00377-3}}.

\bibitem{Witten:1998qj}
Edward Witten.
\newblock {Anti-de Sitter space and holography}.
\newblock {\em Adv. Theor. Math. Phys.}, 2:253--291, 1998.
\newblock \href {http://arxiv.org/abs/hep-th/9802150}
  {\path{arXiv:hep-th/9802150}}, \href
  {http://dx.doi.org/10.4310/ATMP.1998.v2.n2.a2}
  {\path{doi:10.4310/ATMP.1998.v2.n2.a2}}.

\bibitem{Strominger:2001pn}
Andrew Strominger.
\newblock {The dS / CFT correspondence}.
\newblock {\em JHEP}, 10:034, 2001.
\newblock \href {http://arxiv.org/abs/hep-th/0106113}
  {\path{arXiv:hep-th/0106113}}, \href
  {http://dx.doi.org/10.1088/1126-6708/2001/10/034}
  {\path{doi:10.1088/1126-6708/2001/10/034}}.

\bibitem{Witten:2001kn}
Edward Witten.
\newblock {Quantum gravity in de Sitter space}.
\newblock In {\em {Strings 2001: International Conference}}, 6 2001.
\newblock \href {http://arxiv.org/abs/hep-th/0106109}
  {\path{arXiv:hep-th/0106109}}.

\bibitem{Bousso:2000nf}
Raphael Bousso.
\newblock {Positive vacuum energy and the N bound}.
\newblock {\em JHEP}, 11:038, 2000.
\newblock \href {http://arxiv.org/abs/hep-th/0010252}
  {\path{arXiv:hep-th/0010252}}, \href
  {http://dx.doi.org/10.1088/1126-6708/2000/11/038}
  {\path{doi:10.1088/1126-6708/2000/11/038}}.

\bibitem{Spradlin:2001pw}
Marcus Spradlin, Andrew Strominger, and Anastasia Volovich.
\newblock {Les Houches lectures on de Sitter space}.
\newblock In {\em {Les Houches Summer School: Session 76: Euro Summer School on
  Unity of Fundamental Physics: Gravity, Gauge Theory and Strings}}, pages
  423--453, 10 2001.
\newblock \href {http://arxiv.org/abs/hep-th/0110007}
  {\path{arXiv:hep-th/0110007}}.

\bibitem{Anninos:2012qw}
Dionysios Anninos.
\newblock De sitter musings.
\newblock {\em Int. J. Mod. Phys. A}, 27:1230013, 2012.
\newblock \href {http://arxiv.org/abs/1205.3855} {\path{arXiv:1205.3855}},
  \href {http://dx.doi.org/10.1142/S0217751X1230013X}
  {\path{doi:10.1142/S0217751X1230013X}}.

\bibitem{Balasubramanian:2001nb}
Vijay Balasubramanian, Jan de~Boer, and Djordje Minic.
\newblock {Mass, entropy and holography in asymptotically de Sitter spaces}.
\newblock {\em Phys. Rev. D}, 65:123508, 2002.
\newblock \href {http://arxiv.org/abs/hep-th/0110108}
  {\path{arXiv:hep-th/0110108}}, \href
  {http://dx.doi.org/10.1103/PhysRevD.65.123508}
  {\path{doi:10.1103/PhysRevD.65.123508}}.

\bibitem{Anninos:2011ui}
Dionysios Anninos, Thomas Hartman, and Andrew Strominger.
\newblock {Higher Spin Realization of the dS/CFT Correspondence}.
\newblock {\em Class. Quant. Grav.}, 34(1):015009, 2017.
\newblock \href {http://arxiv.org/abs/1108.5735} {\path{arXiv:1108.5735}},
  \href {http://dx.doi.org/10.1088/1361-6382/34/1/015009}
  {\path{doi:10.1088/1361-6382/34/1/015009}}.

\bibitem{Maldacena:2002vr}
Juan~Martin Maldacena.
\newblock {Non-Gaussian features of primordial fluctuations in single field
  inflationary models}.
\newblock {\em JHEP}, 05:013, 2003.
\newblock \href {http://arxiv.org/abs/astro-ph/0210603}
  {\path{arXiv:astro-ph/0210603}}, \href
  {http://dx.doi.org/10.1088/1126-6708/2003/05/013}
  {\path{doi:10.1088/1126-6708/2003/05/013}}.

\bibitem{McFadden:2009fg}
Paul McFadden and Kostas Skenderis.
\newblock {Holography for Cosmology}.
\newblock {\em Phys. Rev. D}, 81:021301, 2010.
\newblock \href {http://arxiv.org/abs/0907.5542} {\path{arXiv:0907.5542}},
  \href {http://dx.doi.org/10.1103/PhysRevD.81.021301}
  {\path{doi:10.1103/PhysRevD.81.021301}}.

\bibitem{Jiang:2025jnk}
Xin Jiang and Haitang Yang.
\newblock {Entanglement entropy of conformal field theory in all dimensions}.
\newblock {\em JHEP}, 01:015, 2026.
\newblock \href {http://arxiv.org/abs/2506.02786} {\path{arXiv:2506.02786}},
  \href {http://dx.doi.org/10.1007/JHEP01(2026)015}
  {\path{doi:10.1007/JHEP01(2026)015}}.

\bibitem{Jiang:2026juf}
Xin Jiang, Peng Wang, and Haitang Yang.
\newblock {Exact Bulk-Boundary Pairs in AdS/CFT}.
\newblock 5 2026.
\newblock \href {http://arxiv.org/abs/2605.15776} {\path{arXiv:2605.15776}}.

\bibitem{Yang:2026mrt}
Haitang Yang.
\newblock {Exact Holographic Kinematics in AdS/CFT}.
\newblock 5 2026.
\newblock \href {http://arxiv.org/abs/2605.21252} {\path{arXiv:2605.21252}}.

\bibitem{deBoer:2015kda}
Jan de~Boer, Michal~P. Heller, Robert~C. Myers, and Yasha Neiman.
\newblock {Holographic de Sitter Geometry from Entanglement in Conformal Field
  Theory}.
\newblock {\em Phys. Rev. Lett.}, 116(6):061602, 2016.
\newblock \href {http://arxiv.org/abs/1509.00113} {\path{arXiv:1509.00113}},
  \href {http://dx.doi.org/10.1103/PhysRevLett.116.061602}
  {\path{doi:10.1103/PhysRevLett.116.061602}}.

\bibitem{deBoer:2016pqk}
Jan de~Boer, Felix~M. Haehl, Michal~P. Heller, and Robert~C. Myers.
\newblock Entanglement, holography and causal diamonds.
\newblock {\em JHEP}, 08:162, 2016.
\newblock \href {http://arxiv.org/abs/1606.03307} {\path{arXiv:1606.03307}},
  \href {http://dx.doi.org/10.1007/JHEP08(2016)162}
  {\path{doi:10.1007/JHEP08(2016)162}}.

\bibitem{Haehl:2025ehf}
Felix~M. Haehl and Kuo-Wei Huang.
\newblock {Virasoro OPE blocks, causal diamonds, and higher-dimensional CFT}.
\newblock {\em JHEP}, 02:030, 2026.
\newblock \href {http://arxiv.org/abs/2505.12456} {\path{arXiv:2505.12456}},
  \href {http://dx.doi.org/10.1007/JHEP02(2026)030}
  {\path{doi:10.1007/JHEP02(2026)030}}.

\bibitem{Nakayama:2015mva}
Yu~Nakayama and Hirosi Ooguri.
\newblock {Bulk Locality and Boundary Creating Operators}.
\newblock {\em JHEP}, 10:114, 2015.
\newblock \href {http://arxiv.org/abs/1507.04130} {\path{arXiv:1507.04130}},
  \href {http://dx.doi.org/10.1007/JHEP10(2015)114}
  {\path{doi:10.1007/JHEP10(2015)114}}.

\bibitem{Wang:2015qfa}
Zhao-Long Wang and Yi~Yan.
\newblock {Bulk Local Operators, Conformal Descendants and Radial
  Quantization}.
\newblock {\em Adv. High Energy Phys.}, 2017:8185690, 2017.
\newblock \href {http://arxiv.org/abs/1507.05550} {\path{arXiv:1507.05550}},
  \href {http://dx.doi.org/10.1155/2017/8185690}
  {\path{doi:10.1155/2017/8185690}}.

\bibitem{Czech:2016xec}
Bartlomiej Czech, Lampros Lamprou, Samuel McCandlish, Benjamin Mosk, and James
  Sully.
\newblock {A Stereoscopic Look into the Bulk}.
\newblock {\em JHEP}, 07:129, 2016.
\newblock \href {http://arxiv.org/abs/1604.03110} {\path{arXiv:1604.03110}},
  \href {http://dx.doi.org/10.1007/JHEP07(2016)129}
  {\path{doi:10.1007/JHEP07(2016)129}}.

\bibitem{Doi:2024nty}
Kazuki Doi, Naoki Ogawa, Kotaro Shinmyo, Yu-ki Suzuki, and Tadashi Takayanagi.
\newblock {Probing de Sitter space using CFT states}.
\newblock {\em JHEP}, 02:093, 2025.
\newblock \href {http://arxiv.org/abs/2405.14237} {\path{arXiv:2405.14237}},
  \href {http://dx.doi.org/10.1007/JHEP02(2025)093}
  {\path{doi:10.1007/JHEP02(2025)093}}.

\bibitem{Ng:2012xp}
Gim~Seng Ng and Andrew Strominger.
\newblock {State/Operator Correspondence in Higher-Spin dS/CFT}.
\newblock {\em Class. Quant. Grav.}, 30:104002, 2013.
\newblock \href {http://arxiv.org/abs/1204.1057} {\path{arXiv:1204.1057}},
  \href {http://dx.doi.org/10.1088/0264-9381/30/10/104002}
  {\path{doi:10.1088/0264-9381/30/10/104002}}.

\bibitem{Bars:1998cs}
Itzhak Bars.
\newblock {Two - time physics}.
\newblock In {\em {22nd International Colloquium on Group Theoretical Methods
  in Physics}}, pages 2--17, 7 1998.
\newblock \href {http://arxiv.org/abs/hep-th/9809034}
  {\path{arXiv:hep-th/9809034}}.

\bibitem{Bars:2000qm}
Itzhak Bars.
\newblock {Survey of two time physics}.
\newblock {\em Class. Quant. Grav.}, 18:3113--3130, 2001.
\newblock \href {http://arxiv.org/abs/hep-th/0008164}
  {\path{arXiv:hep-th/0008164}}, \href
  {http://dx.doi.org/10.1088/0264-9381/18/16/303}
  {\path{doi:10.1088/0264-9381/18/16/303}}.

\bibitem{Bars:2010xi}
Itzhak Bars.
\newblock {Gauge Symmetry in Phase Space, Consequences for Physics and
  Spacetime}.
\newblock {\em Int. J. Mod. Phys. A}, 25:5235--5252, 2010.
\newblock \href {http://arxiv.org/abs/1004.0688} {\path{arXiv:1004.0688}},
  \href {http://dx.doi.org/10.1142/S0217751X10051128}
  {\path{doi:10.1142/S0217751X10051128}}.

\bibitem{Chernikov:1968zm}
N.~A. Chernikov and E.~A. Tagirov.
\newblock Quantum theory of scalar fields in de sitter space-time.
\newblock {\em Annales de l'I.H.P. Physique theorique}, 9:109--141, 1968.

\bibitem{Bunch:1978yq}
T.~S. Bunch and P.~C.~W. Davies.
\newblock Quantum field theory in de sitter space: Renormalization by point
  splitting.
\newblock {\em Proc. Roy. Soc. Lond. A}, 360:117--134, 1978.
\newblock \href {http://dx.doi.org/10.1098/rspa.1978.0060}
  {\path{doi:10.1098/rspa.1978.0060}}.

\bibitem{Mottola:1984ar}
E.~Mottola.
\newblock Particle creation in de sitter space.
\newblock {\em Phys. Rev. D}, 31:754, 1985.
\newblock \href {http://dx.doi.org/10.1103/PhysRevD.31.754}
  {\path{doi:10.1103/PhysRevD.31.754}}.

\bibitem{Allen:1985ux}
Bruce Allen.
\newblock Vacuum states in de sitter space.
\newblock {\em Phys. Rev. D}, 32:3136, 1985.
\newblock \href {http://dx.doi.org/10.1103/PhysRevD.32.3136}
  {\path{doi:10.1103/PhysRevD.32.3136}}.

\bibitem{Collins:2003zv}
Hael Collins.
\newblock A short introduction to the fate of the alpha vacuum.
\newblock 2003.
\newblock \href {http://arxiv.org/abs/hep-th/0312144}
  {\path{arXiv:hep-th/0312144}}.

\bibitem{Jiang:2024hjz}
Xin Jiang, Peng Wang, Houwen Wu, and Haitang Yang.
\newblock {How Einstein{\textquoteright}s equations emerge from CFT2}.
\newblock {\em Phys. Rev. D}, 112(8):L081906, 2025.
\newblock \href {http://arxiv.org/abs/2410.19711} {\path{arXiv:2410.19711}},
  \href {http://dx.doi.org/10.1103/zg5x-34mn} {\path{doi:10.1103/zg5x-34mn}}.

\bibitem{Jiang:2025lkx}
Xin Jiang and Haitang Yang.
\newblock {Derive Einstein equation from CFT entanglement entropy}.
\newblock 10 2025.
\newblock \href {http://arxiv.org/abs/2510.27250} {\path{arXiv:2510.27250}}.

\end{thebibliography}

\newpage

\section{Appendix: AdS as The Scale-Space of  Open-Solid-Torus \label{appendix}}
As a precursor to the present work, Ref. \cite{Yang:2026AdS} 
demonstrated how an AdS QFT can be constructed from a  CFT via 
the Open-Solid-Torus (OST) construction.
This preprint has remained under arXiv moderation for more than one and 
a half months, with no clear timeline for its release. For the sake of self-contained completeness, 
we therefore have to include a concise summary in 
this appendix outlining how the AdS geometry emerges from the OST prescription.

\begin{figure}[h]
\centering
\includegraphics[scale=0.45]{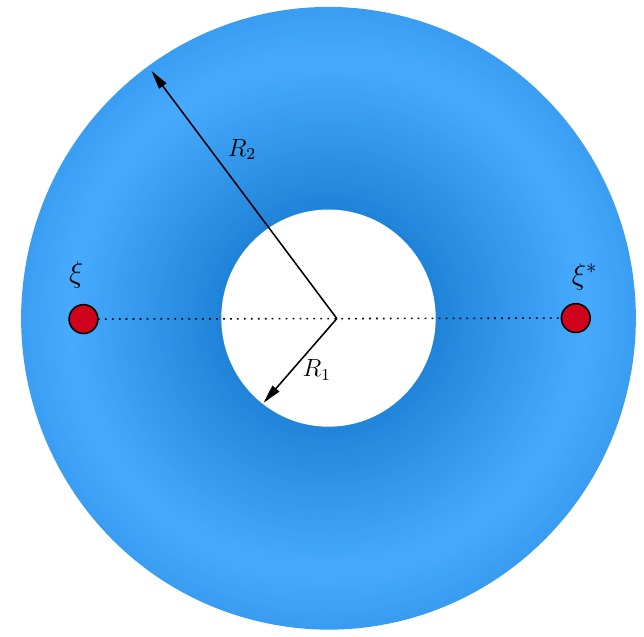}
\caption{The CFT$_d$ lives on the flat symmetric Open Solid Torus $S^1\times\mathbb B^{d-1}$ with Euclidean flat metric.
Every point $\xi$ is paired with a symmetric partner ${\xi^\ast}^\mu = -\xi^\mu$. 
The pairing in a generic asymmetric OST
is obtained by the associated conformal transformation.
\label{fig:solid-torus}}
\end{figure}

Consider a CFT in $d$-dimensional spacetime with flat metric
$ \mathrm d s_E^2=\mathrm d t_E^2+\mathrm d y^2+
\sum_{I=1}^{d-2}\mathrm d x_I^2,$ where Euclidean signature is chosen.
Refer to Fig.~\ref{fig:solid-torus}, let us address
quantum fields  on a symmetric Open-Solid-Torus (OST)  $S^1\times\mathbb B^{d-1}$,
\begin{equation}
\mathcal B_d=\left\{\left(\sqrt{t_E^2+y^2}-\frac{R_2+R_1}{2}\right)^2+
\sum_{I=1}^{d-2}x_I^2<\left(\frac{R_2-R_1}{2}\right)^2\right\},
\end{equation}
with $R_2>R_1>0$ and  coordinate transformation, $t_{\text{E}}=r\sin\theta$,
$y=r\cos\theta$.

\begin{figure}[h]
\centering
\includegraphics[scale=0.8]{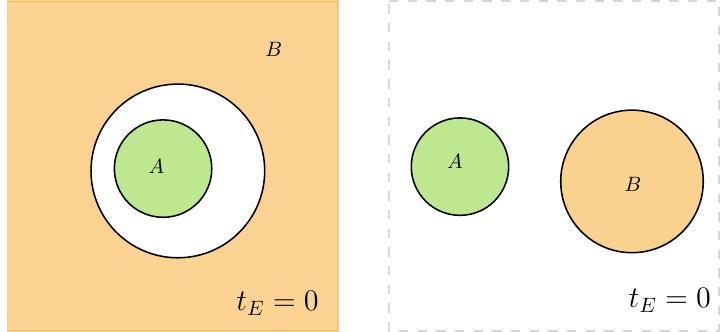}
\caption{$t_E =0$ slice (colored regions)  of the solid torus.
Left panel: the cavity configuration.
Right panel: the juxtaposed configuration.}
\label{fig:two-config}
\end{figure}

Fig.~\ref{fig:solid-torus} is a  symmetric OST.
Fig. \ref{fig:two-config} shows the generic  asymmetric OSTs.
All OSTs  are classified into conformal equivalence classes
by the inversive product,
\begin{equation}
\varrho=\left|\frac{r^{2}+r^{\prime2}-\vert\vec{x}-\vec{x}^{\prime}\vert^{2}}{2rr^{\prime}}\right|,
\label{Appen-eq:inversive product}
\end{equation}
where $\vec x$ and $\vec x'$ are centers and  $r$ and $r'$ are radii of the  two spheres $A$ and $B$ fixing
the OST.

The key is  that points inside an OST are automatically paired.
This is most transparent in the symmetric OST in Fig. \ref{fig:solid-torus}, 
in which each point $\xi^\mu$ is paired with a symmetric partner ${\xi^\ast}^\mu = -\xi^\mu$.
Then we can make a linear parametrization,
\begin{equation}
X^\mu=\frac{\xi^\mu+{\xi^\ast}^\mu}{2},
\qquad
Z=\frac{|\xi^\mu-{\xi^\ast}^\mu|}{2}.
\label{Appen-eq:sysRe}
\end{equation}
The pairing and parametrization in a
generic asymmetric OST are obtained by the associated conformal transformation,
which are complicated nonlinear expressions.

Therefore, the OST leads to a $(d+1)$-dimensional Scale-Space 
\begin{equation}
\boxed{Y^a(\xi) \equiv (X^\mu (\xi),Z(\xi)),\quad \mu=0,\dots, d-1,}
\end{equation}
entirely defined by the $d$-dimensional CFT.

The  geometry of the Scale-Space can be computed by 
using entanglement entropy \cite{Jiang:2024hjz,Jiang:2025lkx} or correlators
\cite{Jiang:2026juf}.
Here, we employ the same ambient embedding  outlined in Eqs. (\ref{eq:ambient-inner})-(\ref{eq:ambient-Poincare}). 
We only need address the symmetric OST. 
Generic asymmetric OSTs follow the same pattern.
Mapping the pairing (\ref{Appen-eq:sysRe}) onto the embedding space, with the unity normalization,
we have
\begin{equation}
V^A(X,Z) = \frac{\big(P^A(\xi)+P^A(\xi^\ast)\big)/2}{\big|P^A(\xi)+P^A(\xi^\ast)\big|/2} 
= \left( \frac{1+X^2 +Z^2}{2Z}, \frac{X^\mu}{Z},  \frac{1-(X^2 +Z^2)}{2Z} \right),\quad V^2=-1.
\label{eq:emb-V}
\end{equation}
Note that this is  a timelike vector, 
rather than a spacelike vector for the ball-moduli in eq. (\ref{eq:ambient-norm}). 
The metric of the the Scale-Space is now induced from the ambient metric:
\begin{equation}
ds_\mathrm{SS}^2 = d V^A d V_A = \frac{dX^\mu dX_\mu + dZ^2}{Z^2},
\label{eq:Metric}
\end{equation}
which is precisely AdS$_{d+1}$. 

After obtaining the Scale-Space metric, following the similar steps in the main text,
we can lift  all the CFT operators to the  Scale-Space with Casimir completion. 
The correlation functions are also lifted.
The source functional and the corresponding 1PI functional are defined accordingly.
The traditional GKPW is recovered as a singular limit.

\end{document}